\documentclass[aps,pre,twocolumn,longbibliography]{revtex4-2}
\usepackage[utf8]{inputenc}
\usepackage{amsmath}
\usepackage{commath}
\usepackage[pdftex]{graphicx}
\usepackage{siunitx}
\usepackage{hyperref}
\usepackage{float}
\usepackage{caption}
\usepackage{subcaption}
\usepackage{comment}
\usepackage{lipsum}
\usepackage{graphicx}

\newcommand{\ext}{\text{ext}}
\newcommand{\fext}{F_{\ext}}

\newcommand{\fv}[1]{\left\langle #1 \right\rangle}
\newcommand{\Ph}{\(P(\tilde{h})\) }

\expandafter\def\expandafter\normalsize\expandafter{%
    \normalsize%
    \setlength\abovedisplayskip{14pt}%
    \setlength\belowdisplayskip{14pt}%
    \setlength\abovedisplayshortskip{14pt}%
    \setlength\belowdisplayshortskip{14pt}%
}

\hypersetup{hidelinks=true}

\begin{document}
\title{Height distribution of elastic interfaces in quenched random media}
\author{Tuuli Sillanpää}\email{tuuli.sillanpaa@tuni.fi} 
\author{Sanni Nousiainen}
\author{Lasse Laurson}\email{lasse.laurson@tuni.fi}
\affiliation{Computational Physics Laboratory, Tampere University, P.O. Box 600, FI-33014 Tampere, Finland}

\begin{abstract}
Elastic interfaces in quenched random media driven by external forces exhibit a continuous depinning phase transition between pinned and moving phases at a critical external force. Recent work [Phys. Rev. Lett. 129, 175701 (2022)] has shown that the distribution of local interface heights at depinning displays negative skewness.
Here, by considering local, long-range and fully-coupled (mean-field) elasticity, we expand on this result by demonstrating the robustness of the negative skewness at depinning when approaching the thermodynamic limit and considering different values of the spring stiffness controlling the avalanche cutoff. 
Additionally, we investigate the evolution of the height distribution as the external force is ramped up from zero, approaching the critical force from below. Starting from a symmetric height distribution at zero force, the distribution initially develops positive skewness increasing with the external force, followed by a steep drop to the negative value characteristic of the critical point as the depinning transition is reached.
 \end{abstract}

\maketitle

\section{Introduction}

A large class of physical systems, including domain walls in ferromagnets~\cite{zapperi1998dynamics} and ferroelectrics~\cite{paruch2005domain}, contact lines in wetting~\cite{joanny1984model}, dislocations~\cite{zapperi2001depinning} and crack fronts~\cite{laurson2013evolution} in disordered solids can be described as driven elastic interfaces in quenched random media. Such systems exhibit critical-like response to slow external driving, manifested as both dynamical signatures like avalanche dynamics with a broad, power-law avalanche size distribution and as interface roughness usually characterized by a single roughness exponent $\zeta$~\cite{wiese2022theory}. These features emerge due to the interplay between quenched disorder, elasticity, and an external driving force, resulting in a zero temperature depinning phase transition between pinned and moving phases of the interface at a critical external force $\fext\ = F_{\mathrm{c}}$~\cite{nattermann1992dynamics,chauve2000creep}. Continuous non-equilibrium depinning phase transitions exhibit scaling and universality~\cite{wiese2022theory}. Here, we focus on three important cases of study 
for linear 1D elastic interfaces in 2D quenched random medium, i.e., interfaces with local, long-range (LR), and fully-coupled [here also referred to as mean-field (MF)] elastic interactions, respectively~\cite{laurson2013evolution,tanguy1998individual,kolton2018critical}, with each case characterized by their own set of critical exponents and universal scaling functions.

Recently, the paradigm of self-affine roughness of elastic interfaces in random media characterized by a single roughness exponent $\zeta$ has been challenged by the observation that the distribution $P(h_i)$ of the local interface heights $h_i$ at criticality exhibits negative skewness~\cite{toivonen2022asymmetric}. This negative skewness originates from a broken symmetry between interface segments above and below the mean interface height due to the direction of the external driving force~\cite{toivonen2022asymmetric}, and has also been independently analyzed by considering the three-point function (see Fig.~31 of Ref.~\cite{wiese2022theory}). As a related consequence, it was found that a spectrum of local, segment-level scaling exponents are needed to fully characterize the rough morphology of the interface~\cite{toivonen2022asymmetric}. More recently, the negative skewness of $P(h_i)$ was demonstrated to play a crucial role in the emergence of the "bump" in the cutoff scaling function $f(x)$ of the distribution of avalanche sizes $s$, $P(s)=s^{-\tau}f(s/s^*)$, where $\tau$ is the avalanche size exponent and $s^*$ 
the cutoff avalanche size~\cite{laurson2024criticality}; the bump has also been predicted by functional renormalization group calculations~\cite{rosso2009avalanche,le2009statistics,le2009size}. These key results emphasize the importance of the skewed $P(h_i)$ to account for fundamental features of depinning criticality of elastic interfaces in random media. Moreover, such ideas have potential applications also in many other driven systems exhibiting avalanche dynamics, e.g., in plastic deformation of crystalline~\cite{berta2025identifying,kurunczi2023avalanches,ispanovity2014avalanches} and amorphous~\cite{jocteur2025protocol,oyama2021unified} materials. However, the study of Ref.~\cite{toivonen2022asymmetric} focused on the critical point only, and no systematic investigation of the dependence of the skewness on the system size was presented. Hence, it is a pertinent question how the skewness of $P(h_i)$ may depend on, e.g., the system size $L$ -- the behavior in the thermodynamic limit $L \rightarrow \infty$ being of particular interest -- and on the value of the external driving force, i.e., the ``control parameter'' of the depinning phase transition. In general, one can study the depinning transition by either controlling directly the driving force or by fixing the mean interface velocity and continuously adjusting the force so that the desired velocity is obtained on the average. The depinning transition then takes place either at the critical value of the force or when taking the limit of the velocity approaching zero from above, i.e., in the so-called quasistatic limit. Both protocols are expected to be compatible, and the choice depends on convenience in the specific case at hand.

Here, by considering local, long-range and fully-coupled (mean-field) elasticity 
we expand on the above-mentioned results, by first studying how the skewness of $P(h_i)$ evolves with the system size $L$ and the spring constant $K$ of the ``driving spring'' in quasistatic constant velocity driving, with $K$ controlling the avalanche cutoff $s^*$. We focus especially on the outcome at the critical point in the thermodynamic limit $L \rightarrow \infty$, $K \rightarrow 0$. Additionally, we consider the evolution of the skewness of $P(h_i)$ with the magnitude of the external driving force $F_\mathrm{ext}$ in force-controlled simulations, by slowly ramping up $F_\mathrm{ext}$ from zero towards $F_\mathrm{ext}=F_\mathrm{c}$. Our results clearly show that the negative skewness of $P(h_i)$ at criticality, probed by the quasistatic constant velocity protocol, is essentially independent of $L$ and $K$, strongly suggesting that the results discussed above remain valid in the thermodynamic limit. Regarding the evolution of the skewness of $P(h_i)$ with $F_\mathrm{ext}$, we find in our force-controlled simulations that the initially symmetric $P(h_i)$ at $F_\mathrm{ext}=0$ first develops positive skewness as $F_\mathrm{ext}$ is ramped up from zero, due to only part of the interface moving forward in avalanches. When approaching $F_\mathrm{ext}=F_\mathrm{c}$ from below, the skewness exhibits a rapid drop to its characteristic negative value at $F_\mathrm{ext}=F_\mathrm{c}$. We argue that this steep drop in skewness of $P(h_i)$ with increasing $F_\mathrm{ext}$ in the immediate vicinity of $F_\mathrm{ext}=F_\mathrm{c}$ is a key morphological signature of the onset of the depinning phase transition from the pinned to the moving phase.

The paper is organized as follows: The next section (Section~\ref{sec2}) presents the numerical methods used in this study, focusing mostly on the relevant models of elastic interfaces driven in quenched random media. This is followed by presenting the results in Section~\ref{sec3}. Section~\ref{sec4} finishes the paper with discussion and conclusions.


\begin{figure}[t]
    \centering
    \includegraphics[width=\columnwidth]{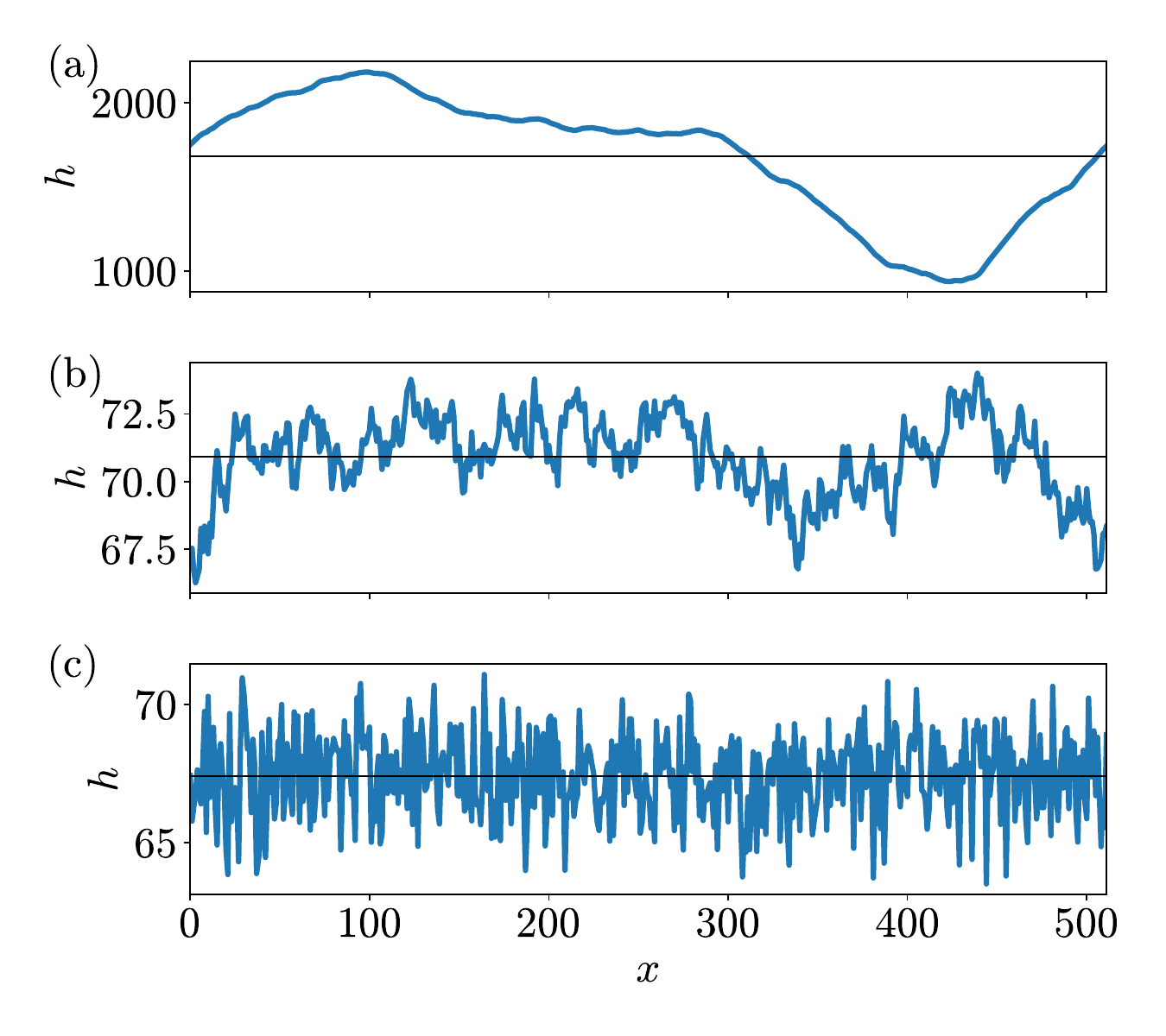}
    \caption{Examples of rough interface configurations at the depinning threshold obtained from the discrete-time models with quasistatic constant velocity driving, with (a) local, (b) long-range and (c) mean-field interactions. The system size is $L=512$ for all three cases. The black horizontal lines indicate the mean interface heights.}
    \label{fig:config}
\end{figure}

\begin{figure*}[t]
    \centering
    \includegraphics[width=0.8\textwidth]{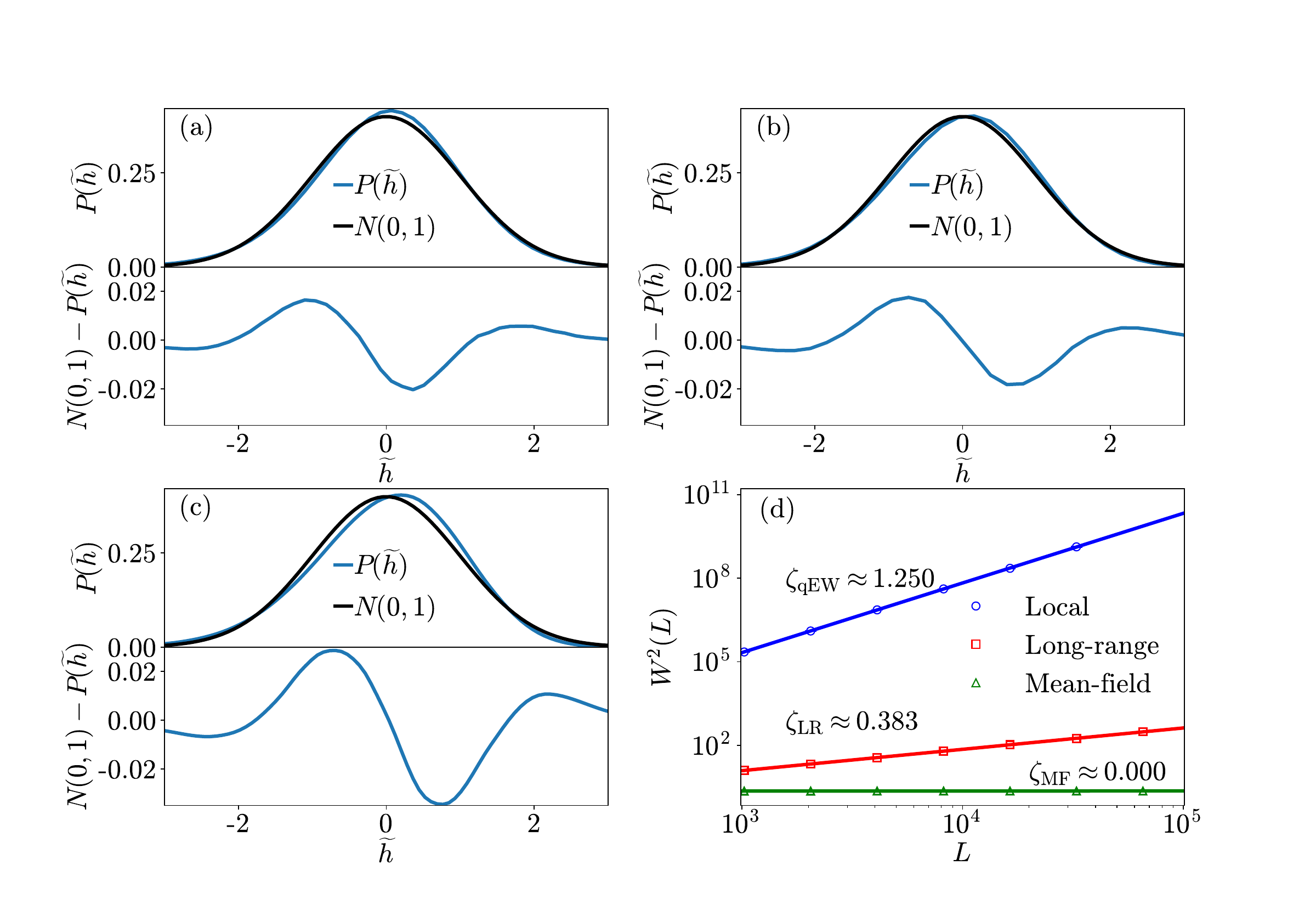}
    \caption[width=\textwidth]{(a) Distributions \(P(\tilde{h})\) of the scaled interface heights for the discrete-time models with quasistatic constant velocity driving (i.e., at depinning) compared to the standard normal distribution \(N(0,1)\) (top) and difference \(N(0,1)-P(\tilde{h})\) (bottom) for the local, long-range and mean-field models in (a), (b) and (c), respectively, indicating negative skewness of \(P(\tilde{h})\). (d) The roughness exponent \(\zeta\) estimated from the scaling of \(W^{2}(L)\) for the different models.}
    \label{fig:dists}
\end{figure*}

\section{Methods}
\label{sec2}

We study numerically models of 1D elastic interfaces driven through 2D quenched random media, considering local, long-range, and fully-coupled (mean-field) elasticity. An instantaneous interface configuration is described by the height function $h(x)$, which we discretize so that $h(x) = h(x_i) = h_i$, where $x_i = i$ (with $i=1, 2, ..., L$, where $L$ is the system size) is an integer-valued coordinate along the $x$ direction (i.e., the average direction of the interface). Periodic boundary conditions are implemented in the $x$ direction. The local total force $F(x_i)$ acting on the interface element \(i\) located at \(x=x_i=i\) is
\begin{equation} \label{eq:total force}
    F(x_i)=F_{\rm el}(x_i)+\eta (x_i,h_i)+F_{\rm ext}.
\end{equation}
The first term on the rhs is the elastic force, which is due to either local, long-range or fully-coupled (mean-field) interactions. Local elasticity is defined as in the discretized quenched Edwards-Wilkinson (qEW) equation~\cite{kardar1986dynamic, alava2002interface, kim2006depinning},
\begin{equation} \label{eq:qEW}
    F_{\rm el,qEW}=\Gamma_0\nabla^2h(x_i)=\Gamma_0(h_{i+1}+h_{i-1}-2h_i),
\end{equation}
with interface stiffness \(\Gamma _0\). Long-range interactions are described by~\cite{gao1989first,bonamy2008crackling,laurson2010avalanches,janicevic2016interevent},
\begin{equation} \label{eq:long}
    F_{\rm el,LR}=\Gamma_0\sum_{j\neq i} \frac{h_j-h_i}{\lvert {x_j-x_i} \rvert ^2}.
\end{equation}
When considering periodic boundary conditions, Eq.~(\ref{eq:long}) becomes~\cite{tanguy1998individual}
\begin{equation}\label{eq:lr}
    F_{\rm el,LR}(x_i)=\Gamma_0\left(\frac{\pi}{L}\right)^2\sum_{j\neq i} \frac{h_j-h_i}{{\sin^2(\frac{x_j-x_i}{L}{\pi})}}.
\end{equation}
The fully-coupled limit of the model is obtained by considering infinite-range interactions, such that~\cite{zapperi1998dynamics,laurson2013evolution,laurson2024criticality}
\begin{equation}\label{eq:mf}
    F_{\rm el,MF}(x_i)=\Gamma_0(\fv{h}-h_i),
\end{equation}
where $\fv{h}=1/L \sum_i h_i$. In Eq.~(\ref{eq:total force}), $\eta$ is uncorrelated Gaussian quenched disorder with mean zero and unit variance, and
$F_\mathrm{ext}$ is the external driving force, pushing the interface in the positive $h$-direction, i.e.,  perpendicular to the average direction of the interface. The stiffness of the interface is set to \(\Gamma_0=0.5\) for all models. System sizes \(L\) considered range from \(L=64\) to \(L=32768=2^{15}\). In what follows, we consider two different dynamics for the model: Discrete-time dynamics to probe the depinning critical point using the quasistatic constant velocity driving protocol, and continuous-time dynamics to study the $F_\mathrm{ext}$-dependence of $P(h_i)$ in force-controlled simulations.

\subsection{Discrete-time dynamics at depinning}
The parallel dynamics of the interface are defined in discrete time $t$ by setting the local velocity
\begin{equation}\label{eq:discrete velocity}
    v(x_i,t)=h(x_i,t+1)-h(x_i,t)=\theta[F(x_i)],
\end{equation}
where \(\theta\) is the Heaviside step function, and $h(x_i)$ are integer-valued discretized interface heights. $\eta(x_i,h_i)$ in Eq.~(\ref{eq:total force}), entering Eq.~(\ref{eq:discrete velocity}) via $F(x_i)$, are uncorrelated Gaussian random numbers with mean zero and unit variance, which are locally updated by drawing a new uncorrelated random number from the Gaussian distribution whenever the local interface height $h_i$ at $x_i=i$ increases by one when the interface moves locally forward [i.e., when $v(x_i,t)=1$], hence meeting new disorder. In order to keep the interface close to the depinning threshold we apply the quasistatic constant velocity driving protocol such that whenever the activity stops [$v(x_i,t)=0$ for all $x_i$], \(\fext\) is increased just enough to make one interface element move [\(F(x_i)>0\) for some \(i\)] and then decreased during the avalanche according to
\begin{equation}\label{eq:decreasing fext}
     \dot{F}_{\rm ext}=-\frac{K}{L} \sum_i v_i(t),
\end{equation}
where the dot denotes the time derivative and \(K\) controls the cutoff of the avalanche size distribution. Physically, $K$ could represent the demagnetizing factor in ferromagnets~\cite{skaugen2019analytical}, or the stiffness of the specimen-machine system in mechanical deformation experiments~\cite{kurunczi2021dislocation}. To study its effect on the skewness, we vary \(K\) in the range from \(0.00008\) to \(0.01024\). Interface configurations \(h(x)\) are stored from the steady state at regular long enough intervals in order to keep the subsequent configurations almost uncorrelated from each other. Depending on the system size and interaction range, the results are averaged over approximately $10^{4}-10^{6}$ line profiles $h(x)$ which are obtained from 10-500 simulations with different disorder realizations.

\begin{figure*}[t]
    \centering
    \includegraphics[width=\textwidth]{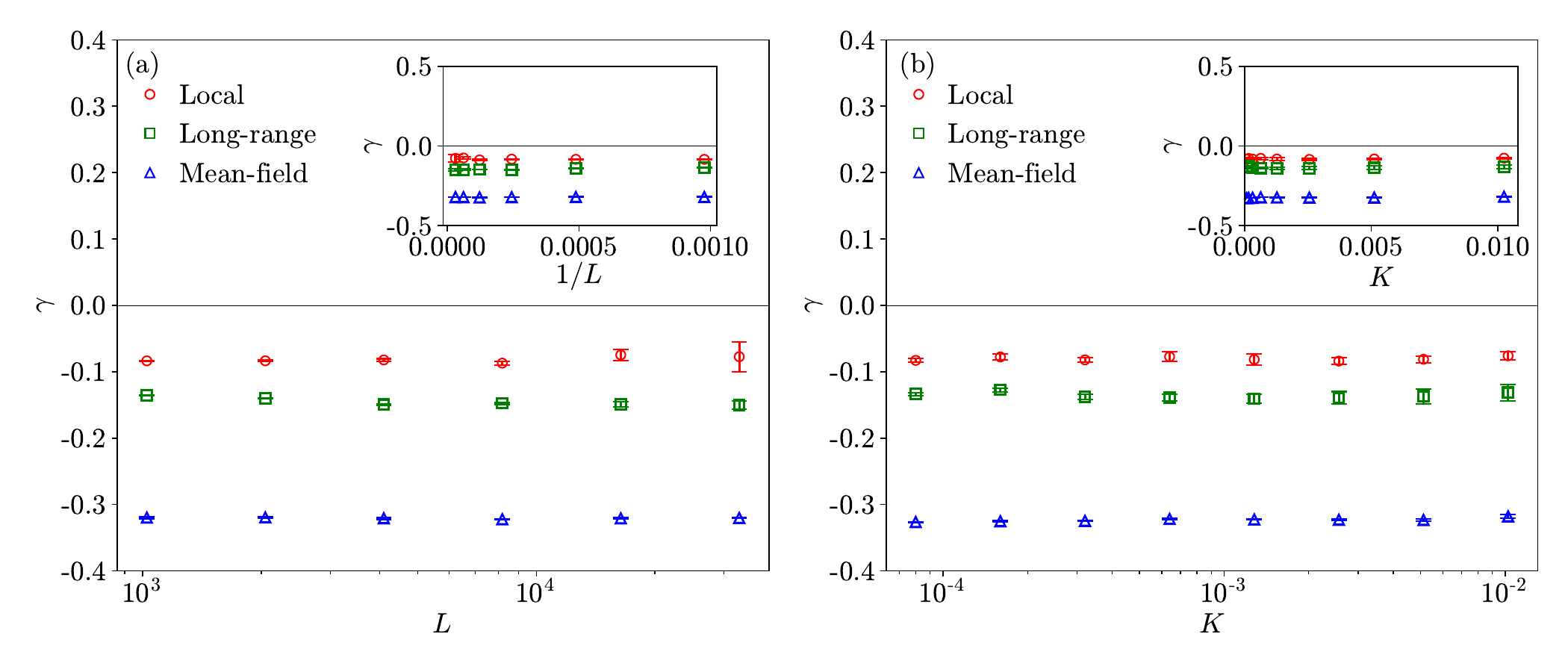}
    \caption{(a) Skewness \(\gamma\) of the distribution \(P(\tilde{h})\) for the different discrete-time models employing the quasistatic constant velocity driving (i.e., at depinning) as a function of the system size \(L\) (main panel) and \(1/L\) (inset) with \(K=0.00625\). (b) Skewness \(\gamma\) as a function of driving spring stiffness \(K\) using a logarithmic (main panel) and linear (inset) $K$-axis with \(L=4096\). Error bars are SEM.}
    \label{fig:LS635}
\end{figure*}

\subsection{Continuous-time dynamics with varying $F_\mathrm{ext}$}

The discrete-time model described above is numerically efficient in sampling the critical point of the depinning transition, but it suffers from the fact that the microscopic dynamics defined by Eq.~(\ref{eq:discrete velocity}) break the symmetry between forward and backward motion as only forward motion is allowed. This becomes an issue in the proximity of $F_\mathrm{ext}=0$, where one expects symmetric behavior, e.g., during the relaxation dynamics from a flat initial configuration. Therefore, to study the dependence of $P(h_i)$ on $F_\mathrm{ext}$, we need to consider a continuous-time variant of the model with a linear mobility law, such that also backwards motion is locally possible. 

To this end, in the simulations with continuous time dynamics we increase the external force \(\fext\) continuously from zero towards the critical force \(\fext=F_\mathrm{c}\) with a slow constant rate $\dot{F}_{\rm ext}=0.00001$. 
The real-valued local interface heights $h_i$ are then updated in parallel according to a linear mobility law,
\begin{equation}\label{eq:continuous loc}
     v(x_i) = F(x_i).
\end{equation}
In practice, for simplicity we use the Euler method to integrate Eq.~(\ref{eq:continuous loc}), $h_i(t+dt)=h_i(t)+F_i(h_i,t)dt$, where $dt$ is the time step. For the continuous-time model, a continuous disorder field $\eta(x_i,h_i)$ is defined for each $x_i$ by first preparing a 2D grid of uncorrelated Gaussian random numbers with zero mean and unit variance, and then employing spline interpolation in the $h$-direction separately for each $x_i$ to obtain a smooth, continuously varying disorder field. Hence, the quenched noise term $\eta$ in Eq.~(\ref{eq:total force}) becomes a force field varying smoothly in space, allowing one to use a simple ODE solver. Interface configurations \(h(x)\) are collected at specific values of \(\fext\) during the continuous force ramp, and the force ramp is repeated several times averaging over the disorder to collect statistics. Depending on the system size and interaction range, we averaged over $10^{3}-10^{5}$ line profiles $h(x)$ from 10-500 simulations with different disorder realizations.

\subsection{Skewness computation}

We calculate the skewness \(\gamma\) of the distribution $P(h_i)$ as an average of the configuration skewnesses determined by the Fisher-Pearson coefficient of skewness
\begin{equation}\label{eq:FP}
    \gamma =\frac{m_3}{m_2^{3/2}},
\end{equation}
where $m_n=\sum_{i=1}^{N} (h_i-\langle h \rangle)^n$ and $N$ is the sample size.
Standard deviation $\sigma$ of the observations is also calculated as an average over the configurations.
Errors are determined by the standard error of the mean, $SEM=\frac{\sigma}{\sqrt N}$. When considering the quasistatic constant velocity driving protocol, even if we store interface configurations in the steady state so that successive configurations are separated by a relatively large displacement, subsequent interface configurations may still be weakly correlated. Hence, when computing the $SEM$-values in that case, instead of the actual sample size, we consider the effective sample size
\begin{equation}\label{eq:ESS}
    N^*=\frac{N}{1+2\sum_{t=1}^{T} \rho_t},
\end{equation}
where we choose $T=10$ and $\rho_t$ is the autocorrelation across configurations
\begin{equation}\label{eq:rho_t}
    \rho_t=\sum_{i=1}^{N-t} \frac{x_i x_{i+t}}{(N-t)\sigma_x^{2}},
\end{equation}
with $x_i=g_i-\langle g \rangle$. $g_i$ is the observable (skewness $\gamma$ or standard deviation $W$) calculated for each configuration, and $\langle g \rangle$ is the mean of $g_i$-values. The choice $T=10$ is motivated by the observation that $\rho_t$ decays quickly to zero and considering larger $T$-values would hence not change $N^*$ apart from statistical noise due to $\rho_t$ oscillating around zero for $t>10$. In general, depending on $\rho_t$, $N^*$ may take values in the range $[1,N]$.



\section{Results}
\label{sec3}

\subsection{Skewness of $P(h_i)$ at the critical point}

Fig. \ref{fig:config} shows example interface profiles \(h(x)\) for all three discrete-time models with \(L=512\) at the critical point, illustrating the qualitative differences between the rough morphologies of the different models. Notice how the local qEW model [Fig. \ref{fig:config}(a)] exhibits only a few very large, almost system-spanning excursions around the mean height (indicated by the black horizontal lines in Fig. \ref{fig:config}), and how these excursions are much more abundant and significantly more limited in size for the long-range model [Fig. \ref{fig:config}(b)]. For the fully-coupled/mean-field model [Fig. \ref{fig:config}(c)], the small local deviations from the mean height display white noise due to the lack of spatial structure in the model. 
The probability distributions \(P(\tilde{h})\) of the scaled interface heights \(\tilde{h}=[h(x)-\langle h \rangle]/\sigma_h\) calculated from an ensemble of interface configurations from the quasistatic constant velocity simulations of the discrete-time models are shown in Fig. \ref{fig:dists}(a), (b), and (c) for the local, long-range, and mean-field model, respectively. In agreement with previous results~\cite{toivonen2022asymmetric}, compared to the standard normal distribution \(N(0,1)\), a long negative tail of \(P(\tilde{h})\), a signature of negative skewness, is observed for all models. As discussed in Ref.~\cite{toivonen2022asymmetric}, this skewness reflects the broken symmetry between interface elements above and below the mean interface height $\langle h \rangle$ due to the direction of the external driving force. Moreover, the scaling of the variance $W^2(L) \equiv \sigma^2_h$ of $P(h_i)$ with $L$, shown in Fig. \ref{fig:dists}(d), allows one to extract the roughness exponents from \(W^{2}(L)\sim L^{2\zeta}\) for the three models. The calculated values ($\zeta=1.25$, 0.38 and 0.0 for the local, long-range, and fully-coupled models, respectively) are very close to the well-known literature values of $\zeta$~\cite{kim2006depinning,rosso2003depinning,duemmer2007depinning,rosso2002roughness}.

We then proceed to analyze the skewness of $P(h_i)$ at the depinning threshold as a function of $L$ and $K$. The skewness of \Ph remains approximately constant as a function of \(L\) [see Fig. \ref{fig:LS635}(a)]. We obtain \(\gamma\approx-0.08\), \(\gamma\approx-0.13\) and \(\gamma\approx-0.32\) independent of $L$ for the local, long-range and fully-coupled models, respectively. Notice that these values are a little smaller in absolute value for the local and long-range models than those reported in Ref.~\cite{toivonen2022asymmetric} for the same $\Gamma_0$; this is due to the calculation procedure employed here where the skewness is taken to be the average of the skewnesses of the individual interface configurations, while in Ref.~\cite{toivonen2022asymmetric} a single skewness value was computed for the whole dataset consisting of a large number of interface configurations. We have checked that the results are otherwise the same irrespective of the averaging procedure, e.g., when it comes to dependence (or lack thereof) of $\gamma$ on $L$ and $K$. Notice also that these values reflect our choice of $\Gamma_0 = 0.5$, and one expects different negative skewness values for other values of $\Gamma_0$, with especially the skewness of the fully-coupled model exhibiting a strong dependence on $\Gamma_0$, while the skewness of the local model is essentially independent of $\Gamma_0$ (see the Supplementary Material of Ref.~\cite{toivonen2022asymmetric}).
Inset of Fig. \ref{fig:LS635}(a) further illustrates that \(\gamma\) does not change in the \(1/L\rightarrow0\) limit, corresponding to \(L\rightarrow\infty\), indicating that the distribution \Ph is negatively skewed at the depinning threshold even in the thermodynamic limit.
Moreover, Fig. \ref{fig:LS635}(b) shows that we obtain the same essentially constant negative values for \(\gamma\) also as a function of \(K\), implying that the skewness of the distribution \Ph is independent of the cutoff $s^* \propto K^{-1/\sigma_k}$ of the avalanche size distribution. 
Hence, we have established that negative skewness of the distribution of local interface heights is a robust feature at the critical point of the depinning transition, such that it is independent of $L$ and $K$ for all the three models considered.


\begin{figure}[t]
    \centering
    \includegraphics[width=\columnwidth]{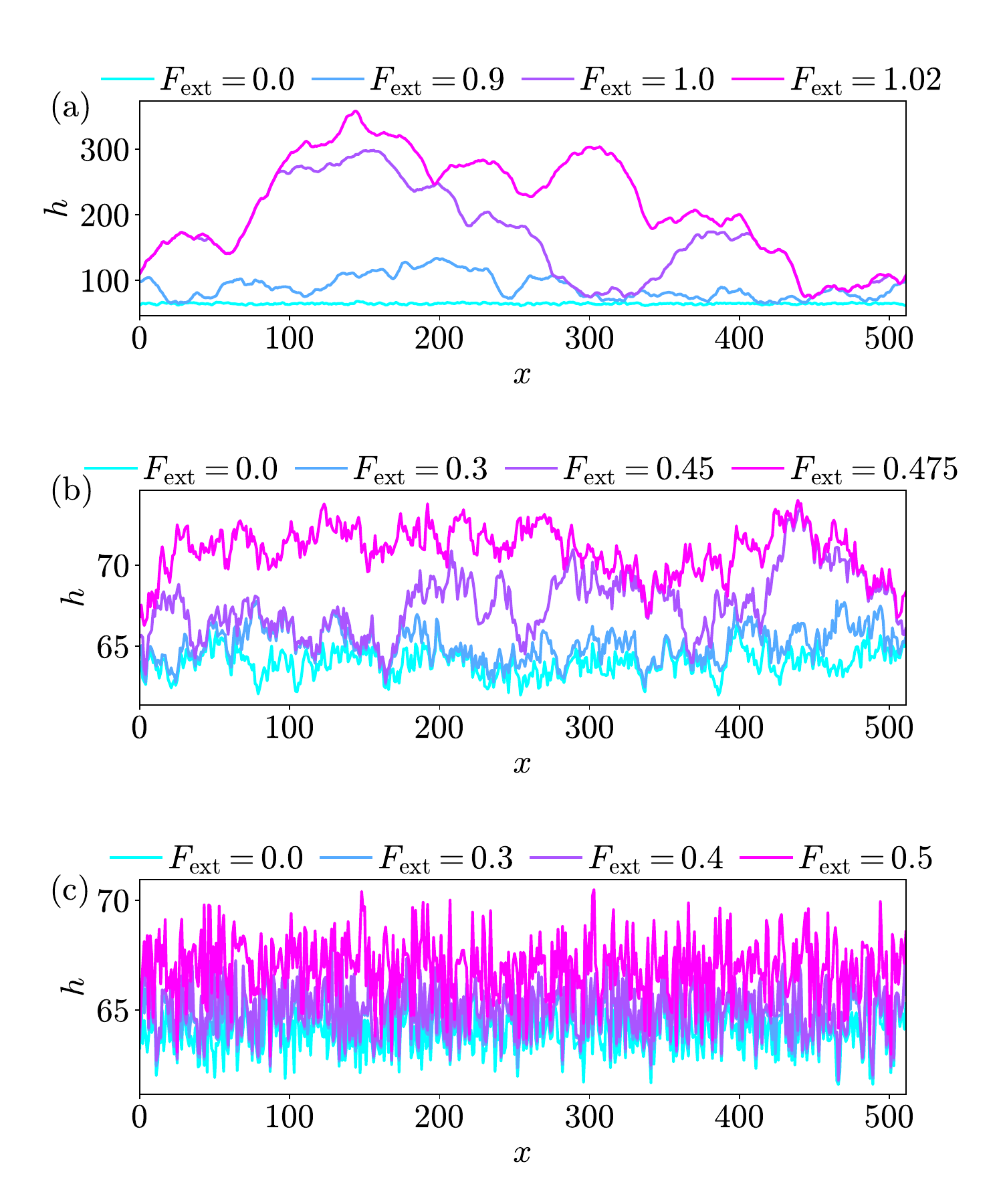}
    \caption{Evolution of the interface height profiles $h(x)$ in the continuous-time models as the external force \(F_{\ext}\) is increased slowly from zero towards $F_\mathrm{c}$ (approximately the largest $F_\mathrm{ext}$-value shown for each model) for (a) the local, (b) the long-range and (c) the mean-field model.}
    \label{fig:forceEvolution}
\end{figure}

\subsection{Skewness of $P(h_i)$ as a function of \(\fext\)}

Next, we proceed to study how the skewness of $P(\tilde{h})$ evolves with \(\fext\) when starting from a relaxed interface configuration of the continuous-time model at \(\fext=0\) and slowly increasing \(\fext\) towards the system size dependent critical value \(\fext=F_\mathrm{c}(L)\). Example interface configurations at different \(\fext\)-values are illustrated in Fig.~\ref{fig:forceEvolution} for the three models (local, long-range and fully-coupled/mean-field). The initial state at \(\fext=0\) is obtained by relaxing a flat initial configuration. As \(\fext\) is slowly ramped up from zero towards the depinning threshold, the interface begins to move via a sequence of momentary avalanches, resulting in roughening of the interface. Initially this implies that only parts of the interface move forward from the relaxed configuration while the rest of the interface stays put. As \(\fext\) reaches \(\fext=F_\mathrm{c}\) (see the interface configurations with the largest \(\fext\) shown in Fig.~\ref{fig:forceEvolution} which are for \(\fext \approx F_\mathrm{c}\)), the entire interface has typically moved away from the initial relaxed configuration, and the rough morphologies of the interfaces qualitatively resemble those shown in Fig.~\ref{fig:config} from the steady state of the quasistatic constant velocity simulations of the corresponding discrete-time models.

\begin{figure}[t]
    \centering
    \includegraphics[width=0.93\columnwidth]{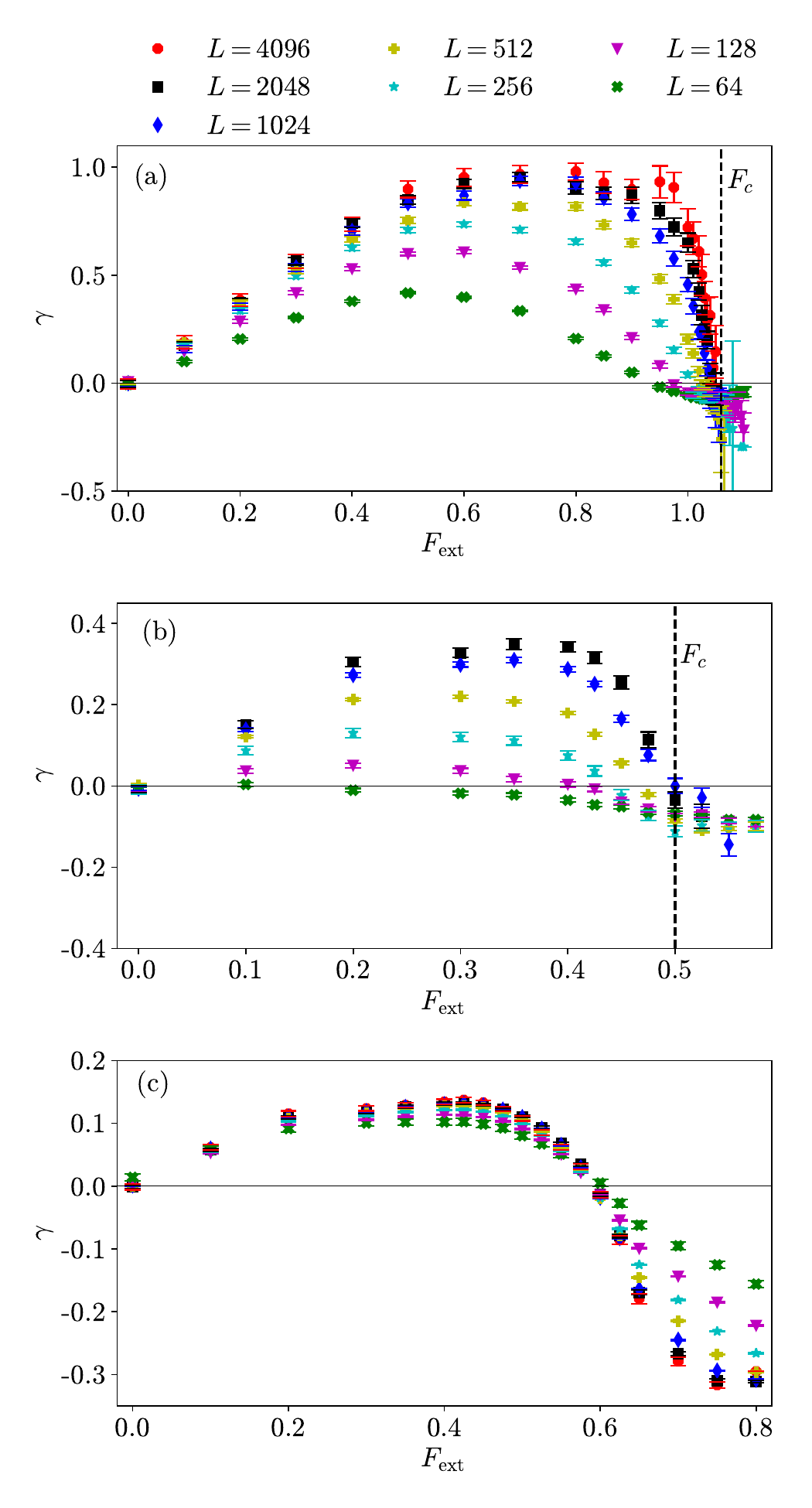}
    \caption{Skewness \(\gamma\) of the distribution \(P(\tilde{h})\) from the continuous-time models for different system sizes \(L\) as a function of the external force \(F_{\rm ext}\) with (a) local, (b) long-range, and (c) mean-field interactions. The vertical dashed lines in (a) and (b) indicate the values of the critical forces $F_\mathrm{c}$ obtained from fits shown in Fig.~\ref{fig:forceStdAll}. Error bars are SEM.}
    \label{fig:forceAll}
\end{figure}
 
 The consequences of the above-discussed features for the skewness $\gamma$ of $P(\tilde{h})$ as a function of \(\fext\) are reported in Fig.~\ref{fig:forceAll}. At \(\fext=0\), $\gamma$ corresponding to the interface configurations relaxed from a flat initial state is zero within statistical fluctuations. This is expected as in the absence of a non-zero \(\fext\) there is nothing that would break the symmetry between interface segments above and below the mean interface height. When \(\fext\) is slowly increased from zero towards the depinning threshold, the roughening of the interface via propagation by a sequence of avalanches implies that some interface elements are located well above the mean interface height while rest of the interface stays put, causing the distribution to become positively skewed as observed in Fig.~\ref{fig:forceAll}.
 As \(\fext\) increases further towards $F_\mathrm{c}$, the avalanches become bigger and hence the interface segments above the mean interface height become larger and propagate further ahead the rest of the interface, resulting in the positive skewness increasing in magnitude (see Fig.~\ref{fig:forceAll}). We note that for the local and long-range models, this effect is more pronounced in larger systems that are able to accommodate larger localized avalanches. On the other hand, the $\gamma$ vs \(\fext\) curve of the fully-coupled mean-field model is to a good approximation independent of $L$ due to the lack of spatial structure in the model, implying that avalanches are not localized in space. Overall, these observations strongly suggest that the skewness trends discussed above will persist in the thermodynamic limit.

As \(\fext\) reaches a value very close to \(\fext=F_\mathrm{c}(L)\), we observe a rapid drop of $\gamma$ from a positive value for \(\fext<F_\mathrm{c}(L)\) to the characteristic negative value at \(\fext \approx F_\mathrm{c}(L)\). The latter are within statistical fluctuations the same in magnitude as the $\gamma$-values observed before when studying the critical point with the discrete-time model with quasistatic constant velocity driving. For the local and long-range models, this effect is again more dramatic in systems of larger size, such that in the limit of a large $L$ the drop of $\gamma$ vs \(\fext\) curve close to \(\fext=F_\mathrm{c}(L)\) seems to asymptotically become vertical. This is especially visible in the case of the local model where already the largest system considered ($L=4096$) exhibits an almost vertical drop of $\gamma$ with \(\fext\) close to \(\fext=F_\mathrm{c}(L)\), see Fig.~\ref{fig:forceAll}. In contrast, for the fully-coupled model the drop of $\gamma$ with increasing \(\fext\) is less dramatic and almost independent of $L$. This drop of $\gamma$ with increasing \(\fext\) is due to the onset of continuous propagation of the interface at \(\fext=F_\mathrm{c}(L)\): For \(\fext<F_\mathrm{c}(L)\), only part of the interface has moved via avalanches from the initial relaxed configuration, resulting in positive skewness of $P(h_i)$ as discussed above, while for \(\fext=F_\mathrm{c}(L)\), most of the interface starts to move, such that the skewness is now controlled by a few of the most strongly pinned parts of the interface which lag behind the rest of the interface, resulting in negative skewness of $P(h_i)$. Hence, we argue that the steep drop of $\gamma$ close to \(\fext=F_\mathrm{c}(L)\) in the large-$L$ limit is a key morphological signature of the onset of the depinning phase transition: Slightly below depinning, the morphology of the rough interface is characterized by a positively skewed height distribution. Upon increasing \(\fext\) the interface morphology suddenly changes as the skewness becomes negative when the critical point is reached.

We also note that the data shown in Fig.~\ref{fig:forceAll} suggest that the maximum of the positive skewness seems to set another characteristic force scale below $F_c$, which depends rather strongly on the system size $L$. This dependence is such that this force scale appears to be pushed towards $F_c$ especially for the local and long-range models as $L \rightarrow \infty$, suggesting that this second force scale is present only in finite systems and it would coincide with $F_c$ in the $L \rightarrow \infty$ limit.

\begin{figure}[t]
    \centering
    \includegraphics[width=0.93\columnwidth]{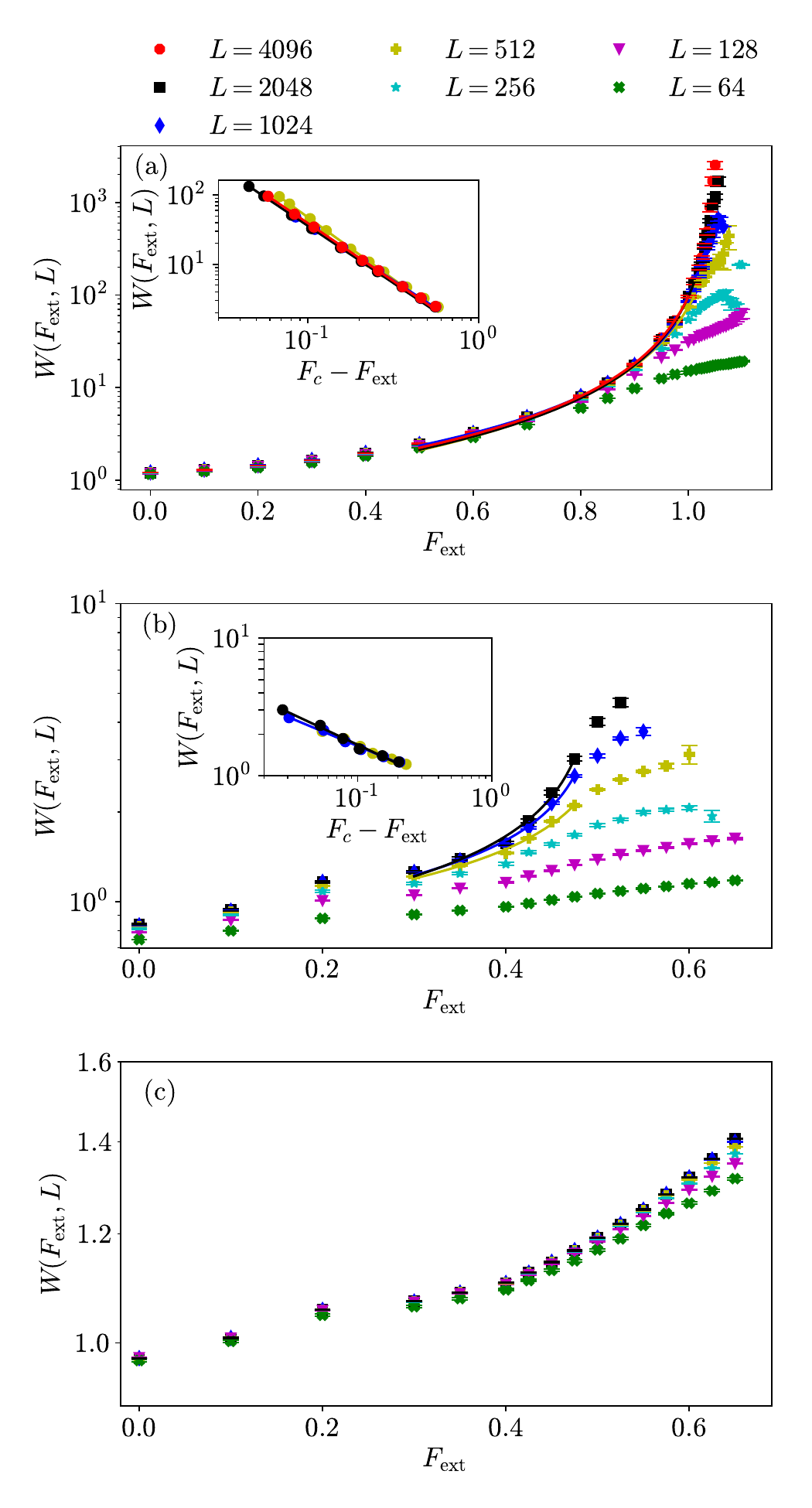}
    \caption{(a) \(W(\fext,L)\) for different \(L\) as a function of \(F_{\rm ext}\) for the local continuous-time model. For the largest $L$'s from \(L=512\) to \(L=4096\), fits of the form of Eq.~(\ref{eq:Wscaling}) are shown with solid lines. Inset presents \(W(\fext,L)\) as a function of \(F_\mathrm{c}-\fext\) where \(F_\mathrm{c}\) is obtained as a fitting parameter from the fit in the main panel (see text). (b) Same as in (a) but for the long-range model, with \(L=2048\) as the maximum system size. (c) \(W(L)\) as a function of \(\fext\) for the mean-field model. Error bars are SEM.}
    \label{fig:forceStdAll}
\end{figure}

Finally, in addition to the skewness \(\gamma\) we also examine the evolution of the width (standard deviation) \(W(F_\mathrm{ext},L)\equiv \sigma_h\) of $P(h_i)$ as a function of \(\fext\) for different system sizes $L$. As \(\fext\) is ramped up towards $F_\mathrm{c}$, the lateral correlation length (i.e., the lateral size of the largest avalanches) \(\xi\) diverges as $\xi \sim [F_\mathrm{c}(L)-\fext]^{-\nu}$. Since the interface roughens due to localized avalanches, in an infinite system the width $W(F_\mathrm{ext})$ is expected to scale as
\begin{equation}
\label{eq:Wscaling}
W(F_\mathrm{ext}) \sim \xi^\zeta \sim (F_\mathrm{c}-\fext)^{-\nu\zeta}.
\end{equation}
For systems with a finite $L$, the divergence of Eq.~(\ref{eq:Wscaling}) is terminated when $\xi$ reaches $L$. Fig.~\ref{fig:forceStdAll} shows \(W(F_\mathrm{ext},L)\) as a function of \(\fext\) for different system sizes $L$, showing also fits of the form of Eq.~(\ref{eq:Wscaling}) to the large-$L$ data close to \(\fext = F_\mathrm{c}(L)\) for the local and long-range models, resulting in independent estimates of the $F_\mathrm{c}(L)$-values for those models as fitting parameters. The fits yield $F_\mathrm{c}=1.06\pm 0.002$ and $F_\mathrm{c}=0.50\pm 0.01$ for the largest system sizes of the local and long-range models, respectively, shown in Figs.~\ref{fig:forceAll}(a) and (b) as vertical dashed black lines. This allows us to verify that the steep drop of $\gamma$ in Fig.~\ref{fig:forceAll} indeed coincides with the $L$-dependent critical force $F_\mathrm{c}(L)$ for the local and long-range models. We observe clear signatures of divergence of $W$ at $\fext = F_\mathrm{c}(L)$ for the local and long-range models in Fig.~\ref{fig:forceStdAll}, with $\nu\zeta = 1.67 \pm 0.04$ and $\nu\zeta = 0.46 \pm 0.09$ for the local and long-range models, respectively, with the latter measured from a rather narrow scaling range. These values can be compared with the literature values of $\nu=1.333$ and $\zeta=1.250$ for the local qEW model~\cite{ferrero2013nonsteady}, resulting in $\nu \zeta = 1.67$, and $\nu=1.625$ and $\zeta=0.385$ for the long-range model~\cite{duemmer2007depinning}, giving $\nu \zeta = 0.63$. However, no such divergence is present in the fully-coupled model since $\zeta=0$ [see Eq.~(\ref{eq:Wscaling})], with $W(F_\mathrm{ext})$ varying only by a factor of $\sim 1.4$ almost independent of $L$ in the entire range of $F_\mathrm{ext}$-values considered [Fig.~\ref{fig:forceStdAll}(c)].

\section{Conclusions}
\label{sec4}
To sum up, we have demonstrated that the earlier results~\cite{toivonen2022asymmetric} showing that the distributions $P(h_i)$ of the local interface heights of one-dimensional interfaces with elastic interactions of different ranges exhibit negative skewness at the depinning threshold persists in the thermodynamic limit. Given the independence of $\gamma$ on $L$ and $K$ it is tempting to argue that it should be possible to derive this result also analytically, e.g., in the fully-coupled limit. Moreover, we have shown how the skewness $\gamma$ of $P(h_i)$ evolves with \(\fext\), from $\gamma=0$ at \(\fext=0\) to a positive $\gamma$ when \(\fext\) is increased from zero, followed by a rapid drop of $\gamma$ to its characteristic negative value at \(\fext = F_\mathrm{c}\). These findings result in two immediate conclusions: (i) A non-zero skewness $\gamma$ of $P(h_i)$ -- positive or negative depending on \(\fext\) -- is a robust feature of elastic interfaces driven through quenched random media, observable even in the thermodynamics limit, and (ii) the rapid, almost vertical drop of $\gamma$ from a positive value for \(\fext<F_\mathrm{c}\) towards the negative $\gamma$ at depinning can be considered to be a key morphological signature of the onset of the depinning transition. We expect that these results will qualitatively carry over to higher-dimensional elastic interfaces. This expectation is supported by our results for the fully-coupled model which can viewed as being effectively a very high-dimensional system.

The negative skewness at depinning -- the robustness of which we have demonstrated here -- has already been shown to play a crucial role in the emergence of the "bump" in the avalanche size distribution of finite-dimensional systems~\cite{laurson2024criticality}. Hence, it is tempting to speculate that also other non-trivial features of finite-dimensional depinning criticality, absent in the mean-field limit, such as the asymmetry of the average avalanche shape~\cite{laurson2013evolution} and/or anticorrelations between subsequent avalanches~\cite{le2020correlations}, could be accounted for by similar mechanisms. Such ideas offer interesting possibilities for future work. Finally, while our study has focused on the specific case of numerical models of depinning of elastic interfaces in quenched random media, the results presented here could find applications both in the relevant experimental systems exhibiting criticality due to depinning, ranging from domain walls in ferromagnetic thin films~\cite{albornoz2021domain} to planar crack fronts~\cite{santucci2010fracture,janicevic2016interevent}, as well as in the context of a wide range of other driven systems displaying criticality and avalanche dynamics, including elasto-plastic models~\cite{jocteur2025protocol}, sheared glasses~\cite{oyama2021unified}, dislocation avalanches~\cite{berta2025identifying, kurunczi2023avalanches,ispanovity2014avalanches}, and possibly even earthquakes~\cite{de2016statistical}. Exploring the properties of the interface height distribution - including its possible skewness and the contribution of the latter to the emergence of the bump in the avalanche distribution - in such experimental systems provides an interesting future research direction.

\begin{acknowledgments}
  The authors acknowledge the computational resources provided by CSC (IT Center for Science, Finland) and TCSC (Tampere Center for Scientific Computing, Finland). This work was supported by Research Council of Finland (grant number 338955).  
\end{acknowledgments}

\section*{Data availability}

The data that support the findings of this article are openly available~\cite{data}.


%

\end{document}